# Precision Magnetometers for Aerospace Applications

James S. Bennett,[1,*] Brian E. Vyhnalek,[2,*] Hamish Greenall,[1] Elizabeth M. Bridge,[1] Fernando Gotardo,[1] Stefan Forstner,[1] Glen I. Harris,[1] Félix A. Miranda,[2,†] and Warwick P. Bowen[1,‡]

[1]*School of Mathematics and Physics, The University of Queensland, St Lucia, Brisbane Queensland 4072, Australia*
[2]*NASA Glenn Research Center, Cleveland, Ohio 44135, The United States of America*

Aerospace technologies are crucial for modern civilization; space-based infrastructure underpins weather forecasting, communications, terrestrial navigation and logistics, planetary observations, solar monitoring, and other indispensable capabilities. Extraplanetary exploration – including orbital surveys and (more recently) roving, flying, or submersible unmanned vehicles – is also a key scientific and technological frontier, believed by many to be paramount to the long-term survival and prosperity of humanity. All of these aerospace applications require reliable control of the craft and the ability to record high-precision measurements of physical quantities. Magnetometers deliver on both of these aspects, and have been vital to the success of numerous missions. In this review paper, we provide an introduction to the relevant instruments and their applications. We consider past and present magnetometers, their proven aerospace applications, and emerging uses. We then look to the future, reviewing recent progress in magnetometer technology. We particularly focus on magnetometers that use optical readout, including atomic magnetometers, magnetometers based on quantum defects in diamond, and optomechanical magnetometers. These optical magnetometers offer a combination of field sensitivity, size, weight, and power consumption that allows them to reach performance regimes that are inaccessible with existing techniques. This promises to enable new applications in areas ranging from unmanned vehicles to navigation and exploration.

## I. INTRODUCTION

Magnetometers are a key component in space exploration missions, particularly in those concerning the study of the Earth from space, as well as the study of the planets in our solar system. The information gathered from these instruments has been of great benefit in increasing our understanding of the composition and evolution of the Earth [1–5], other planets [6–9], and the interplanetary (heliospheric) magnetic field [10, 11]. They are also widely used in technical aerospace applications; for instance, allowing attitude determination [12] and magnetic geological surveying [13, 14]. Extensive overviews of space-based magnetometers have been previously performed by Acuña [15] in 2002, Díaz-Michelena [16] in 2009, and Balogh [17] in 2010, detailing the design, operation, and calibration of magnetometers flown from the Mariner missions of the early 1960s to the Lunar Prospector and Mars Global Surveyor missions of the turn of the century. This review is intended to provide an updated synopsis of aerospace magnetometry, including both extraplanetary applications and those in Earth atmosphere and orbit, as well as emerging technologies and applications.

A particular focus of the review is on emerging magnetometer technologies that use optical readout [18–20], their performance characteristics, and their potential aerospace applications. This is motivated in part by the exponential growth in the use of unmanned aerial vehicles, together with proposals to use magnetometer-equipped drones for extraplanetary exploration [21–23]. The optical magnetometers considered in this review include atomic magnetometers (*e.g.*, [18]), magnetometers based on quantum defects in diamond (*e.g.*, [19]), and optomechanical magnetometers (*e.g.*, [20]). While each of these kinds of magnetometer have quite different characteristics, in general, a key attraction has been that they offer exquisite sensitivity without requiring cryogenic cooling. In recent years they have also experienced rapid miniaturization, with a concomitant reduction in power consumption. This combination of attributes holds promise for new aerospace applications both on Earth and in extraplanetary missions.

* These authors contributed equally.
† felix.a.miranda@nasa.gov
‡ w.bowen@uq.edu.au

## II. EXISTING APPLICATIONS

### A. Interplanetary science missions

Precise magnetic field measurements are critical to the fulfilment of the objectives of many planetary, solar, and interplanetary science missions. Careful measurements of the magnetic fields associated with celestial bodies help the scientific community to better understand and familiarize itself with the laws of space physics at play in the evolution of planets and the solar system. Thus, magnetometers are essential for science mission applications, and space exploration – one of the paramount goals of humankind – as a whole.

Magnetometers have been used primarily for field mapping and characterization [7, 8, 15, 17, 24–26], but also for the study of planetary atmospheres and their climatic evolution due to solar wind interactions – both in-orbit and from the Martian surface [6]; as well as for indirect detection of liquid water – a critical element for the existence of Earth-like life beyond our planet [27].

Below, we provide some examples of relatively recent, high-visibility missions featuring space magnetometers. Table I summarizes the various spacecraft magnetometers' key specifications. Notably, fluxgate magnetometers (FGMs) stand out as the tool of choice, due to their long-proven performance and reliability in the space environment, as well as their ability to comply with stringent requirements (*e.g.*, weight and power consumption) associated with space missions. However, missions requiring exploration of planets/celestial bodies with extreme environments (*i.e.*, high temperatures and/or high radiation) such as those exhibited by Venus, Europa, Enceladus, *etc.*; landing and exploring planetary surfaces (*e.g.*, rovers); as well as missions requiring multiple observation platforms (*e.g.*, small satellite constellations and swarm platforms), may require magnetometers with sensing, configuration, and form factors different from FGMs.

TABLE I: Summary of various planetary and interplanetary spacecraft magnetometer specifications (FGM = fluxgate magnetometer, VHM = vector helium magnetometer, SHM = scalar helium magnetometer).

| Mission | Magnetometer | Dynamic Range (nT) | Resolution (pT) | Mass (kg) | Power (W) |
|---|---|---|---|---|---|
| GOES-1–3 | FGM (biaxial) | 50–400 | - | - | - |
| GOES-4–7 | FGM (biaxial) | ±400 | 200 | - | - |
| GOES-I–M | FGM (triaxial) | ±1000 | 100 | - | - |
| GOES-N–P | FGM (triaxial) | ±512 | 30 | - | - |
| GOES-R | FGM (triaxial) | ±512 | 16 | 2.5 | 4 |
| MAVEN | FGM (triaxial) | ±512 | 15 | - | $> 1$ |
| | | ±2048 | 62 | | - |
| Cassini | FGM (triaxial) | ±40 | 4.9 | 0.44 (FGM) | 7.5 (sleep) |
| | +V/SHM | ±400 | 48.8 | 0.71 (V/SHM) | 11.31 (FGM+VHM) |
| | | $\pm 10,000$ | 1200 | - | 12.63 (FGM+SHM) |
| Juno | FGM (triaxial) | ±1600 (nominal) | 48 | 5 | $> 4.5$ |
| | | $\pm 1,638,400$ (largest) | 5000 | | |

Fluxgate magnetometers [28–33] consist of a drive coil and a sense coil wrapped around a magnetically permeable core. A strong alternating current applied to the drive coil induces an alternating magnetic field in the core, which periodically drives the core into saturation. When there is no background magnetic field the sense current matches the drive current; however the presence of an external magnetic field acts to bias the saturation of the core in one direction, causing an imbalance between the drive and sense currents which is proportional to the magnitude of the external magnetic field. These magnetometers are sensitive to the direction of the external magnetic field and are therefore classed as vector magnetometers. There are many variations on this basic design, including double-core devices that null the sense current in the absence of an external field. This technology provides a magnetic field sensitivity of around 10 pT/ $Hz^{1/2}$ , a DC (direct current, *i.e.*, zero frequency) magnetic field resolution of around 5 pT and a spatial resolution of about 10 mm [29, 32, 33]. The sensitivity of fluxgate magnetometry is limited by the Barkhausen noise from the core and $1/f$ noise at low frequencies [15].

#### *1. Mars Atmosphere and Volatile Evolution (MAVEN)*

The MAVEN mission, part of NASA's Scout program, was launched to Mars on November 18, 2013, and entered into orbit around the red planet on September 21, 2014. Among the primary goals of the mission was to study the role of atmospheric escape in changing the climate of Mars through time. Other objectives of the mission were to assess the Martian upper atmosphere, ionosphere, and interactions with the solar wind, as well as to determine the escape rates of neutral gases and ions, and collect data that will determine the ratios of stable isotopes to better understand the evolution of Mars' atmosphere [24].

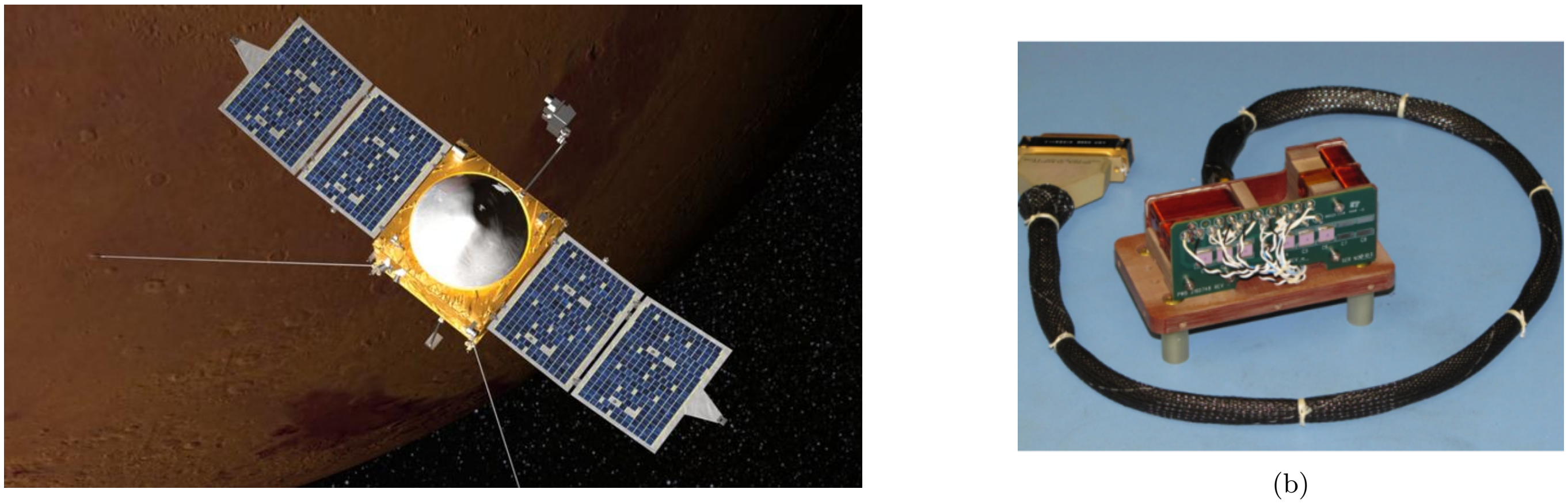

(a) (b)

FIG. 1: (a) Illustration of the MAVEN spacecraft in orbit over Mars. The magnetometer "boomlets" are located at both ends of the solar array panels. (b) MAVEN magnetometer sensor assembly [25]. Credits: NASA/Goddard Space Flight Center.

To facilitate these studies, MAVEN was equipped with a payload of multiple scientific instruments (the "Particles and Fields Package"), including a pair of ring-core FGMs [25]. Drawing upon the heritage of the Mars Global Surveyor mission [15], the MAVEN magnetometers were mounted on "boomlets" at either end of the deployable solar array panels, approximately 5.6 m from the body center, rather than on a dedicated magnetometer boom (Figure 1a). For the Martian field environment the magnetometers have two operating dynamic range modes, $\pm 512$ nT and $\pm 2048$ nT, with digital resolution of 0.015 nT and 0.062 nT, respectively. Additionally, the magnetometer sensors have a high dynamic range mode ($65,536$ nT at 2.0 nT resolution), used for testing in the Earth field environment without requirements for magnetic shielding. A detailed overview of the design, calibration procedures, and performance is given in [25]. The MAVEN Magnetic Fields Investigation plays an important role in understanding how solar wind interactions – including plasma wave formation and structures – lead to atmospheric escape. A picture of the MAVEN magnetometer assembly is shown in Figure 1b.

### *2. Cassini*

The Cassini–Huygens mission, a U.S.–European space mission to Saturn, was launched on October 15, 1997, with the goal of detailed spatio-temporal monitoring of physical processes within the Saturnian system environment, especially in relation to Titan. The orbiter continued to return various science data until 2017, when its fuel supply was exhausted. In particular, magnetometer measurements were made of the internal planetary magnetic field; three-dimensional magnetospheric mapping was performed; the interplay between the magnetosphere and the ionosphere was investigated; and electromagnetic interactions between Saturn, its moons, rings, and the surrounding plasma were observed.

The Cassini magnetometer was a dual system comprised of both a three-axis FGM (three perpendicular ring-core FGMs) and a vector helium magnetometer (VHM), with an additional scalar helium magnetometer (SHM) mode for precise *in situ* absolute calibration of the FGM [8]. As is typical of most spacecraft, the FGM and V/SHM were mounted on a magnetometer boom or "mag boom", as shown in Figure 2. In this case the mag boom was 11 m long, with the V/SHM sensor mounted on the end and the FGM mounted halfway. This configuration allowed for more effective deconvolution of stray magnetic fields associated with the spacecraft from the intended observations. As discussed in [8], the FGM featured four operating ranges spanning $\pm 40$ nT to $\pm 44,000$ nT at resolutions of 4.9 pT and 5.4 nT, respectively, where the largest range was primarily intended for ground testing within the Earth's field. In vector mode, the V/SHM was capable of 3.9 pT resolution across a $\pm 32$ nT dynamic range, and 31.2 pT resolution at $\pm 256$ nT; in scalar mode it had a single range of 256-–$16,384$ nT at 36 pT resolution.

Helium magnetometers [34–38] are often used as secondary magnetometers for calibration of FGMs, which are susceptible to long-term drift. Typically V/SHMs have lower size, weight, and power (SWaP) requirements than FGMs. They have a low operation bandwidth and are generally used for DC measurements. Helium-4 atoms are optically pumped into their $2^3\mathrm{S}_1$ metastable state which contains three Zeeman sub-levels. A radio frequency (RF) source is used to drive the transition between the Zeeman sub-levels, the resonant frequency of which is determined by the background magnetic field $B_0$, through the relationship $f_{\mathrm{RF}} = \gamma_{^4\mathrm{He}} B_0$, where $\gamma_{^4\mathrm{He}}$ is the gyromagnetic ratio $\approx 28$ GHz/T. The amplitude of the resonance signal can be amplified using a population stirring technique where atoms are selectively pumped from metastable Zeeman sub-levels to the $2^3\mathrm{P}_0$ state and subsequently decay back to the metastable state for increased interaction with the incident RF field [38].

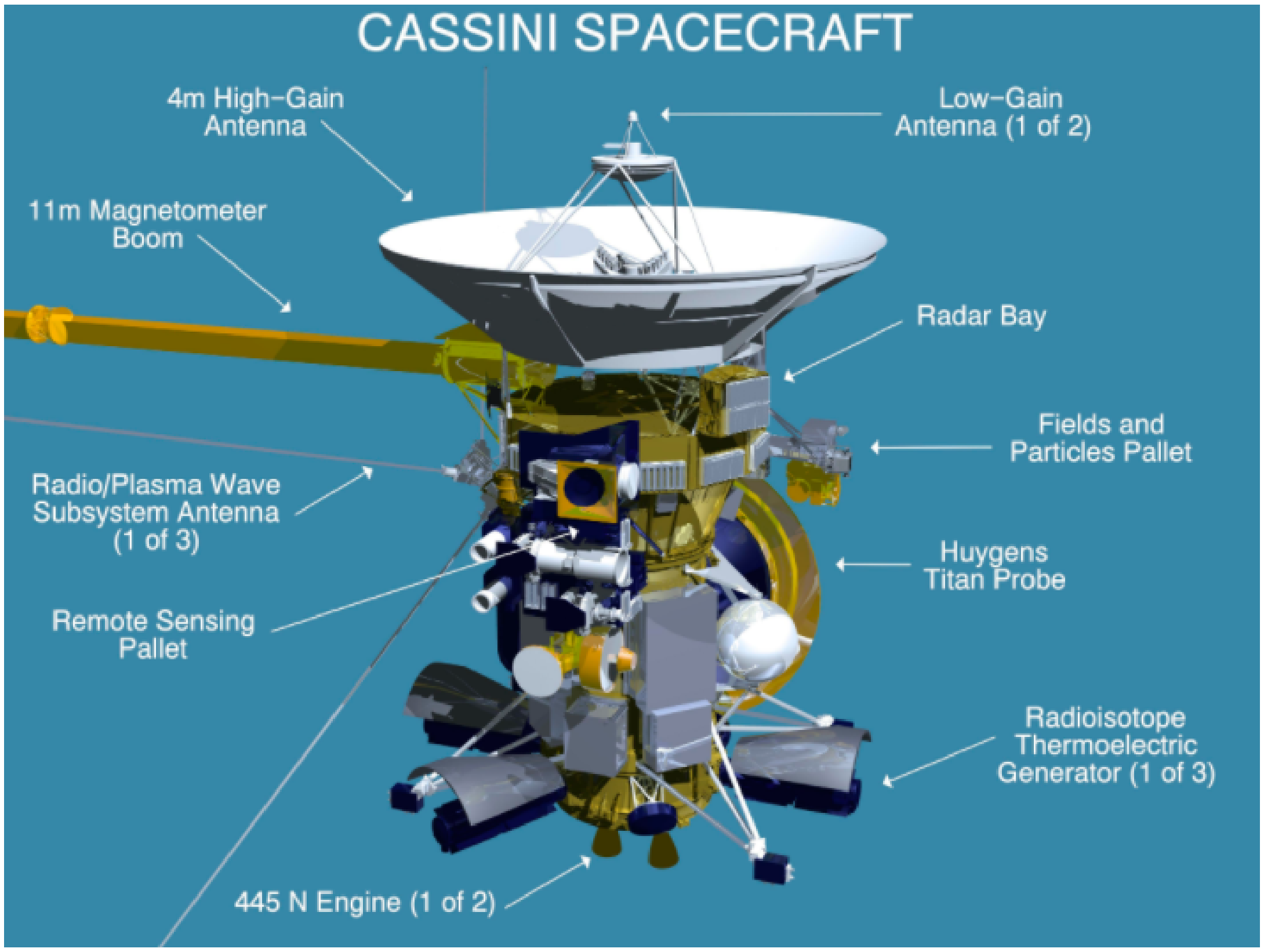

FIG. 2: Diagram of the Cassini spacecraft. Credit: NASA.

#### *3. Juno*

The Juno spacecraft, in orbit around Jupiter since 2016, has the primary mission goals of characterizing Jupiter's planetary magnetic field and magnetosphere. Juno's magnetometers have been used for three-dimensional mapping of Jupiter's magnetic environment, and play an important role in the investigation of the formation and evolution of Jupiter, particularly by allowing scientists to study how the planet's powerful magnetic field (20,000 times stronger than Earth's) is generated [39]. Similar to previous missions, the Juno spacecraft's magnetic field instrumentation utilizes two independent triaxial ring-core FGM sensors, along with co-located non-magnetic imaging sensors (*i.e.*, star trackers), to provide accurate attitude information near the point of magnetic field measurement [7]. The FGMs and star trackers were mounted on vibration-isolated carbon–silicon-carbide platforms on a 4 m boom, nominally 11 m from the spacecraft body. In terms of sensitivity, the magnetometers are capable of six different ranges, extending from a minimum range of $\pm 1600$ nT to a maximum range of $\pm 1,638,400$ nT, with 0.0488 nT resolution in the nominally most sensitive range [7].

### B. Future interplanetary missions

Magnetometers form an integral part of planetary missions being planned or currently under development for future space exploration. While some upcoming high-profile missions – such as NASA's Psyche Discovery mission or the European Space Agency's (ESA) JUpiter ICy moons Explorer (JUICE) – will continue to use improved heritage instrumentation, particularly the widely-used fluxgate magnetometer, there is also an emerging set of potential applications in view of the trend towards smaller platforms and probes (*e.g.*, CubeSats, NanoSats, PocketQubes, *etc.*), in addition to rovers and rotorcraft.

The Psyche Discovery mission aims to study the 16-Psyche asteroid, an asteroid orbiting the sun between Mars and Jupiter, and unique in that it is made almost entirely of nickel-iron metal, unlike the rocky, icy, or gas-covered worlds explored by all other previous space missions. Magnetometry plays a significant role in the mission; the first objective of the mission is to detect and measure a magnetic field, which would confirm that Psyche is the core of a planetesimal [9]. Typical of past interplanetary missions, the Psyche magnetometer consists of two identical fluxgate sensors in a gradiometer configuration located at the middle and outer end of a mag boom. Drawing heritage from the Magnetospheric Multiscale Mission [40], the Psyche magnetometers have two selectable dynamic ranges of $\pm 1000$ and $\pm 10^5$ nT, with resolutions of $\pm 0.1$ and $\pm 10$ pT respectively.

The JUpiter ICy moons Explorer (JUICE), a mission being developed by the European Space Agency (ESA), will have a payload consisting of ten state-of-the-art instruments to carry remote sensing and geophysical studies of the Jovian system. The JUICE spacecraft is scheduled for launch in June 2022, and is set to arrive in orbit around Jupiter in 2030. There, it will perform continuous observations of Jupiter's atmosphere and magnetosphere over a 2.5-year period [41]. Among the instruments of the payload is a magnetometer intended for the characterization of the Jovian magnetic field, its interaction with the internal magnetic field of Ganymede, and the study of the subsurface oceans in the icy moons. The magnetometer is of the fluxgate type, using fluxgate inbound and outbound sensors mounted on a boom [41].

The Europa Clipper Mission being developed by NASA, which will be launched in the 2020's (specific launch date is not yet declared), will conduct studies of Jupiter's moon Europa to determine if the moon could harbor the necessary conditions for the existence of life. Nine scientific instruments will comprise the Europa payload, including cameras, spectrometers, ice-penetration radars, and a triaxial fluxgate magnetometer, among others. The magnetometer will be used to measure the strength and direction of Europa's magnetic field, allowing scientists to determine the depth and salinity of its ocean [42].

Further, CubeSats (satellites built at the scale of 10 cm cubed) continue to gain traction as a suitable platform for breaking new ground in planetary science and exploration. For example, a CubeSat-based distributed fluxgate magnetometer network has been proposed for characterizing Europa's deep ocean [43].

Alongside orbital surveys, a suite of unmanned rovers with terrestrial, atmospheric, and/or oceanic capabilities have been proposed for investigations of extraplanetary bodies, particularly the large moons of Jupiter and Saturn. Airborne extraterrestrial vehicles, as first demonstrated by NASA's Ingenuity flights on Mars [44], have excellent potential for targeted planetary operations. For example, NASA's Dragonfly mission – due for launch in 2026 and projected to arrive at Titan in 2034 – will study the moon's atmospheric and surface properties, along with prebiotic chemistry in its subsurface oceans [21]. Magnetometers are not included in Dragonfly's payload due to size and weight restrictions, highlighting the need for miniaturized and efficient magnetometers for extraterrestrial drones. Other proposed drone missions, such as those submitted to the ESA's 'Voyage 2050' long-term planning process, include magnetometer-equipped missions to both Enceladus [23] and Titan [22] designed to launch within the next thirty years.

### C. Applications in Earth atmosphere and orbit

Magnetometers also serve a critical role in aerospace applications in Earth's atmosphere and orbit, ranging from attitude determination in satellites to geomagnetic surveys using unmanned aerial vehicles (UAVs). Here, we provide an updated synopsis of Earth-based aerospace magnetometry while highlighting the advantages and limitations of existing magnetometers.

#### *1. Geostationary Operational Environmental Satellites (GOES)*

Near-Earth satellites are important platforms for the collection of magnetic data, both for wide-scale geological and military observations [103–107], plus geomagnetic and magnetospheric monitoring. The GOES – part of a series of satellites of the National Oceanic and Atmospheric Administration (NOAA) that have been in operation since the mid 1970s – are a key example of the latter. Their payloads have included magnetometers to measure the Earth's magnetic field, primarily to provide information about geomagnetic storms, energetic particle measurements, and magnetospheric and ionospheric effects. These measurements are particularly important for the characterization of ionospheric scintillation affecting high-precision location measurements with GPS (Global Positioning System) [1], as well as effects on the electric power grid [2], high-frequency radio communications in the 1–30 MHz range [3], and also satellites in low-Earth orbit (LEO), which can experience extra atmospheric drag when solar activity is high [108]. Additionally, the GOES magnetometer data have also been used in real-time support of rocket launch decisions [109].

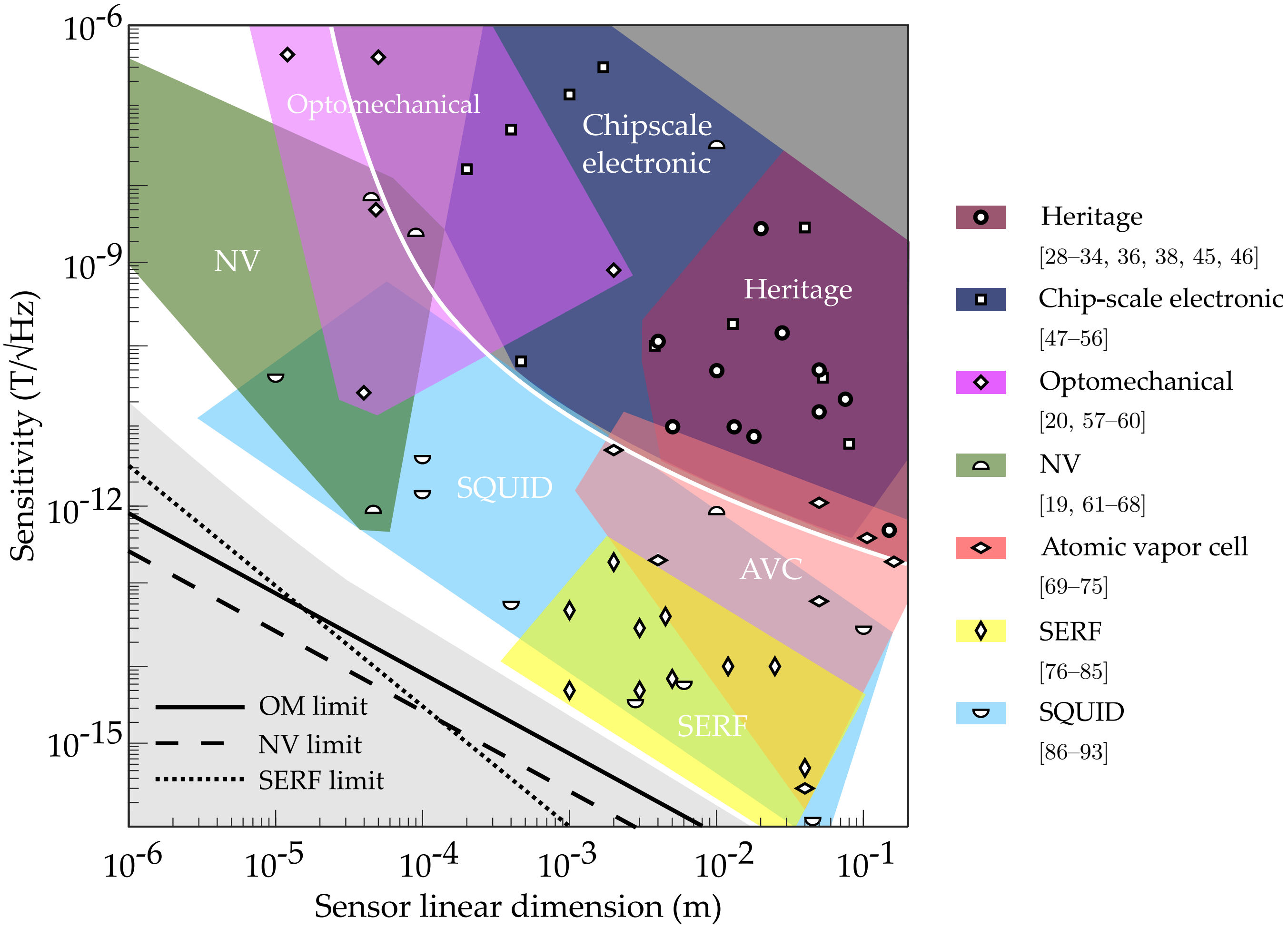

FIG. 3: Sensitivities for available and emerging magnetometers. The colored regions indicate typical parameter regimes for each category of device, with icons showing representative examples from the scientific literature. The dark gray region (bordered by a white line) contains the majority of traditional magnetometers (fluxgate, search coil, proton precession, Overhauser) and chip-scale magnetometers with electrical readout (magnetoresistive, Lorentz-force-actuated MEMS, *etc.*). The light gray region shows the parameter regime that has not yet been explored. Approximate performance limits to some magnetometers are shown as diagonal lines; the thermal limit to optomechanical sensing (solid line, [94]; see also [95, 96]), the atom shot noise limit to SERF sensing (dotted, [97]), and the NV quantum projection noise limit (dashed, [65]).

The initial GOES series – *i.e.*, GOES-1,-2, and -3 (1975—1978) – featured biaxial, closed-loop fluxgate magnetometers (these feature a feedback loop that nulls the external field at the sensor's location). These FGMs were deployed on booms approximately 6.1 m long, with one sensor aligned parallel to the spacecraft spin axis and the other perpendicular, with a sensing range from 50–400 nT. The GOES-4,-5,-6, and -7 (1980–1987) satellites were equipped with spinning twin-fluxgate magnetometers, mounted on 3 m booms, and had a range of ±400 nT with 0.2 nT resolution. Extending the capabilities of the GOES 1–7 spacecraft, the GOES-NEXT series (GOES-I(8) through GOES-M(12)), were launched between 1994 and 2001. This series of spacecraft used two redundant triaxial FGMs, with an increased range of ±1000 nT at a resolution of 0.1 nT. In this case the electronics were located inside the body, with the two magnetometers mounted on 3 m deployable booms. The following installments, GOES-N,-O,-P (13–15), had FGMs of reduced dynamic range, ±512 nT, in favor of a 2× improved resolution of 0.03 nT, and were mounted in a gradient configuration on 8.5 m booms [110]. Finally, the most recent in the set are the GOES-R series (GOES-R/S/T/U) with GOES-R and -S having been launched in 2016 and 2018, respectively. The magnetometers featured here are similar to the GOES-N series triaxial-FGM configurations, but with improved resolution on the order of 0.016 nT [111]. Figure 6 shows an artistic rendition of the GOES-R spacecraft, illustrating the location of the FGM.

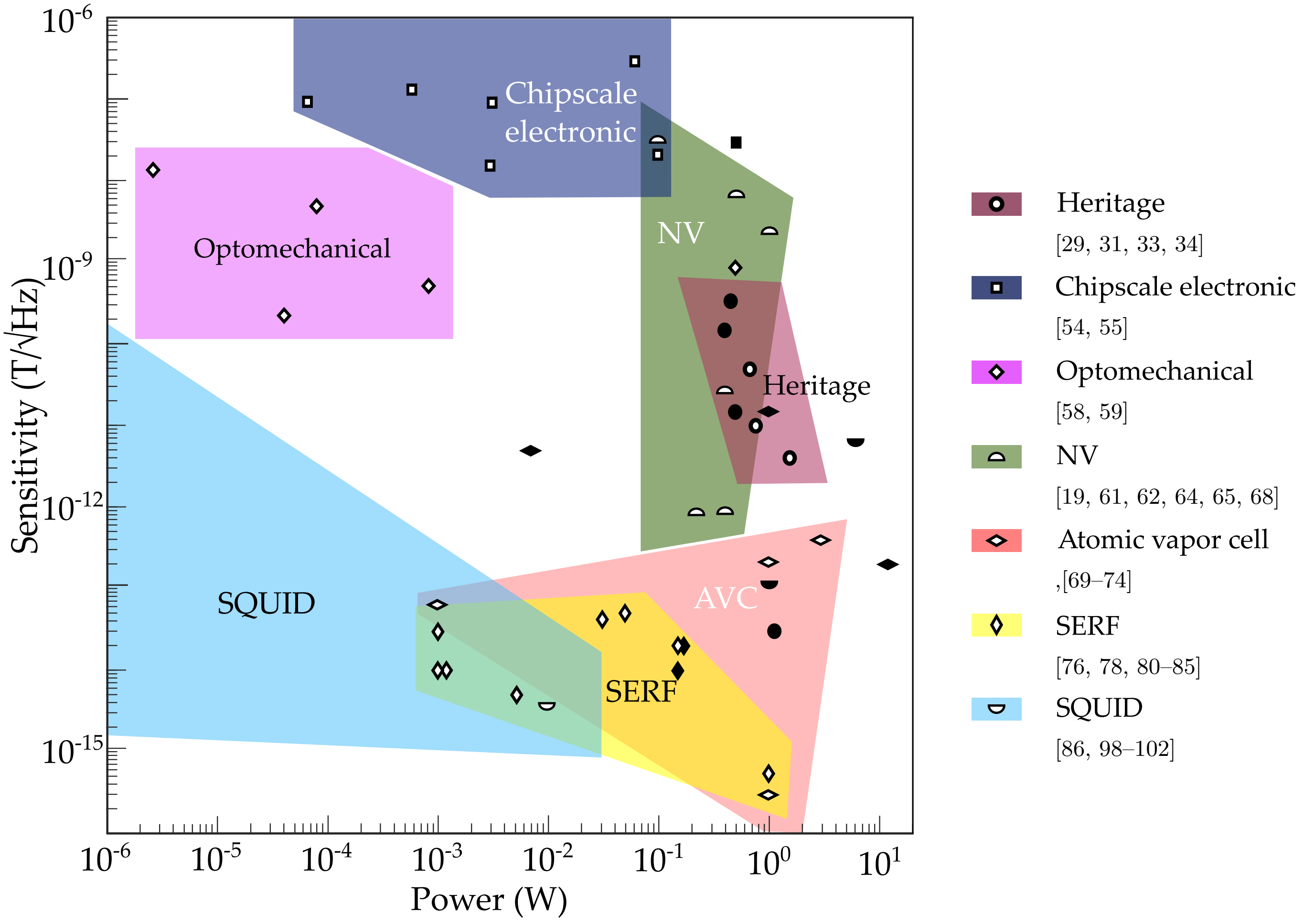

FIG. 4: Sensor power requirements for available and emerging magnetometers, not including support systems such as cryostats (except as indicated below). Colored regions indicate typical parameter regimes for different varieties of magnetometer. Representative examples from the scientific literature are given as 'open' (white with black border) icons. Where available, the total power use of packaged devices (including support systems, *etc.*) are shown as 'solid' black icons. Note that SQUID magnetometers can have extremely low sensor power dissipation ($\sim$ 10 fW) due to their superconducting nature, hence the blue region extends beyond the left border of the plotted region; however, their total power use is typically large due to cryogenic requirements.

Similar magnetometer-equipped satellite networks have been proposed to supplement crucial RADAR-based early warning systems for dangerous tectonic activity [4, 5, 11, 112, 113] through the detection of magnetic anomalies prior to earthquakes (*e.g.*, GESS, Global Earthquake Satellite System).

### 2. *Magnetometers onboard micro- and nanosatellites*

Magnetometers are conventionally used onboard satellites as part of the attitude determination system for low-earth orbit satellites [12]. However, magnetometers cannot usually obtain three-axis attitude information with only a three-axis magnetometer, and the measurement is distorted by magnetized objects and current loops on board the satellite itself. Hence, these systems include other sensors that can measure the satellite's motion with respect to celestial bodies [114].

These additional sensors are too bulky and power-consuming to be used in micro- and nanosatellites, such as CubeSats, which consequently have to rely on attitude determination by magnetometer only. As these small satellites are starting to be applied to more sophisticated objectives, such as remote-sensing and astronomy missions, precise attitude determination is becoming a requirement [114]. The task is additionally complicated by the typically large magnetic moment of satellites with small inertia, which can then cause magnetic bias noise due to the interaction of the earth's magnetic field with the magnetic moment of the satellite [115].

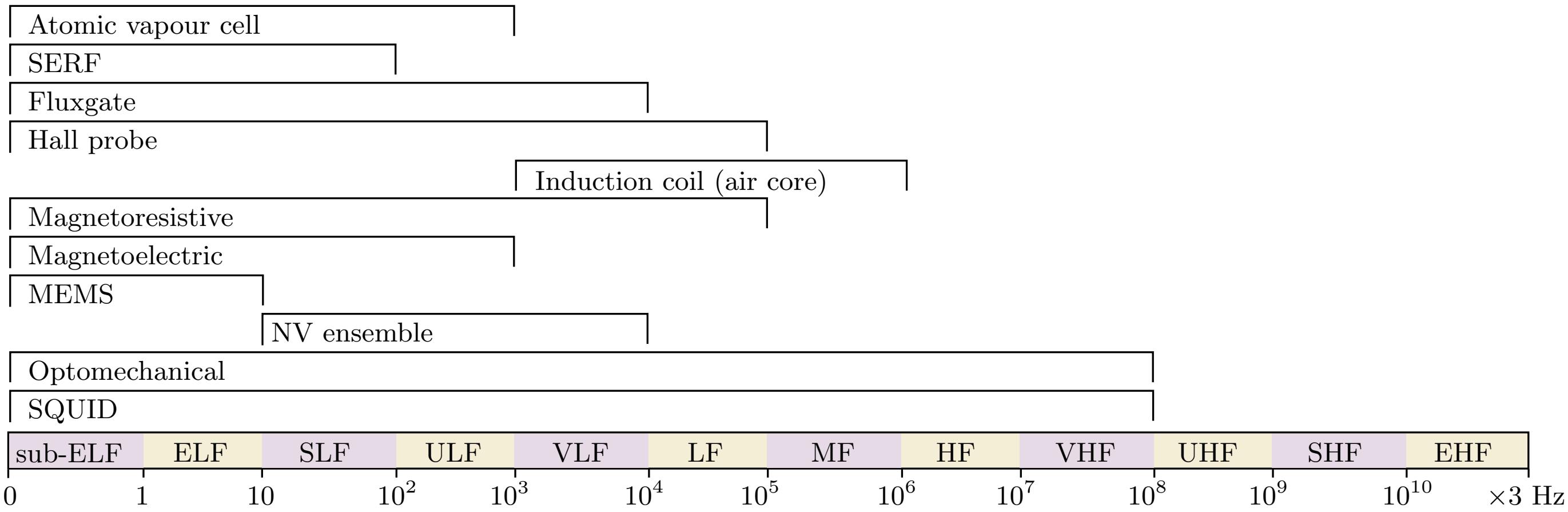

FIG. 5: Operating frequencies for different varieties of magnetometer (grouped by International Telecommunication Union frequency designations).

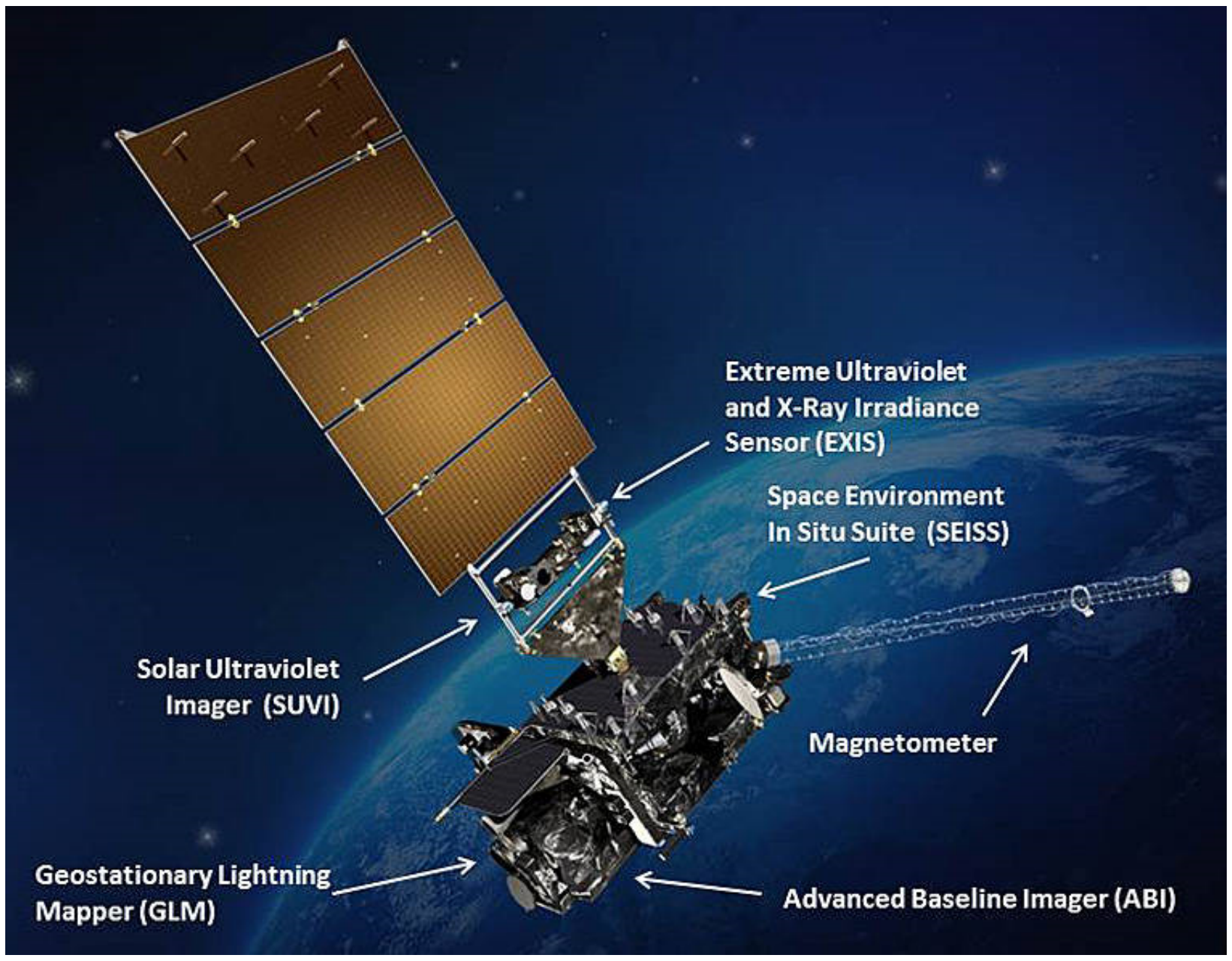

FIG. 6: GOES-R spacecraft (2016). Credit: NASA

A key challenge is therefore to compensate for this magnetic bias noise. This could be achieved by estimating the interaction with the earth's magnetic field using a gyroscope and a Kalman filter [114]. The other major challenge is to achieve full three-axis attitude determination using a magnetometer only. This can, in principle, be achieved by comparing magnetic field readings to an accurate model of the earth's magnetic field. However, this method is computationally expensive [12]. Finally, disturbances onboard the satellite itself could be accounted for by either very careful calibration [116], or by using several magnetometers in different parts of the satellite, which would require further reduction of SWaP.

Considerable work on further SWaP reduction of fluxgate magnetometers has been spearheaded by Todd Bonalsky, Efthyia Zesta, *et al.* from NASA Goddard Space Flight Center [117, 118]. FGMs of significantly reduced SWaP have been developed and deployed on the Dellingr spacecraft launched in 2017 and the Scintillation Prediction Observations Research Task (SPORT) CubeSat that is expected to be launched from the International Space Station in 2021–2022.

NASA's Gateway platform, an orbital outpost, which is intended to be positioned near the Moon as a stepping stone to Mars, will utilize these miniaturized FGMs as part of its space weather monitoring instrument suite, Gateway HERMES (Heliophysics Environmental and Radiation Measurement Experiment). The magnetometers on Gateway HERMES will allow NASA to study the solar winds and the Earth's magnetotail for the purposes of understanding and forecasting solar weather events that will affect astronauts and instruments operating on or around the Moon. The FGMs will be placed on the end of a boom, far away from the Gateway's power and propulsion module. Two magneto-inductive sensors, which have significantly lower SWaP than the FGMs, will be mounted on the Gateway HERMES platform to detect and subtract magnetic noise generated by the power and propulsion module.

Magneto-inductive sensors [45, 46] contain a solenoidal-geometry coil wrapped around a high-permeability magnetic core that forms the inductive element of an LR relaxation oscillation circuit. The effective inductance of the coil is proportional to the magnitude of the magnetic field parallel to the axial direction of the coil. The oscillation frequency of the circuit will vary with the magnetic field at the coil. Commercially available magneto-inductive sensors, such as the PNI RM3100, use comparison with an internal clock to measure the oscillation period of the circuit, and hence the magnitude of the magnetic field. Such sensors are small ($\sim 15$ mm$^3$), have a low operating power ($\sim 0.1$ W), a resolution of around 20 nT, a dynamic range of -800 μT to +800 μT, and a sample rate of $> 400$ Hz.

NASA's Goddard team are also working on developing self-calibrating hybrid devices to overcome the drift experienced by fluxgate magnetometers. These hybrid devices contain a vector fluxgate magnetometer paired with a scalar atomic magnetometer. Their small SWaP makes them suitable for deployment in constellation-type missions where multiple CubeSats simultaneously gather multi-point observations [117].

An alternative to FGMs onboard micro satellites are magnetoresistive magnetometers. These are based on either giant magnetoresistance (GMR) or anisotropic magnetoresistance (AMR). GMR is an effect observed in thin films comprised of sandwiched ferromagnetic and diamagnetic ('non-magnetic') layers (such as Cu). In the presence of a magnetic field, the magnetic moments of the two ferromagnetic layers become aligned and the interlayer resistance decreases drastically [119]. AMR makes use of permalloy (Ni 80%, Fe 20%) that has electrical resistivity that varies as a function of the strength and orientation of the external magnetic field [119]. These techniques have been reported to achieve sensitivity of about 1 nT/ Hz$^{1/2}$ at micrometer scale resolution and under ambient operating conditions; thus, they have seen diverse applications as sensors in biomedicine [120], consumer electronic products such as smart phones [121], and as precision sensors in aerospace applications for low-field magnetic sensing. While AMR magnetometers have historically exhibited hysteresis and stability issues [15], Brown *et al.* have reported on the development of a compact, dual-sensor vector AMR magnetometer for applications on very small spacecraft [122]. The instrument, called MAGIC (MAGnetometer from Imperial College), exhibits sensitivities of 3 nT in a 0–10 Hz band within a measurement range of $\pm 57,500$ nT, at a total mass of only 104 grams, and power consumption in the range of 0.14–0.5 W (depending on the mode of operation). These very low SWaP requirements make magnetoresistive magnetometers suitable for applications in attitude orbit control systems of small satellites – they have already been launched in the TRIO-CINEMA CubeSat space weather mission [123] – as well as planetary landers. Further discussion of AMR/GMR magnetometers (along with microelectromechanical MEMS magnetometers [53–56], which will not be discussed in detail here) can be found in [16].

### *3. Navigation in the Earth's atmosphere*

Magnetometers have been used as a part of airplanes' navigation systems for many years to provide heading information [124, 125]. Historically, observations of the local geomagnetic field have been performed using ground- or aircraft-based proton-precession magnetometers [124–127]. A sample of hydrogen-rich material (typically kerosene) is polarised by the application of a magnetic field; when the field is turned off, the protons precess around the ambient geomagnetic field at a frequency proportional to the field strength. This is detected with an induction coil. Scalar and vector operation is possible [126]. Aerospace applications of proton-precession magnetometers are primarily hindered by their large power consumption. This is addressed by Overhauser magnetometers: built around the same precession phenomena, but leveraging the Overhauser effect [128] to efficiently generate nuclear magnetic polarisation through RF pumping. The resulting sensitivity boost and reduction in SWaP has even allowed Overhauser magnetometers to be flown aboard satellite missions, *e.g.*, the Danish Ørsted satellite (sensitivity $\sim 20$ pT/ Hz$^{1/2}$ , 3 W of power consumption, 1 kg mass) [106, 129, 130].

In recent years, it has been proposed to obtain precise position information by measuring local magnetic field variations and overlapping them with a detailed map of the Earth's magnetic field [131]. This proposal relies on the unique local variations of the Earth's magnetic field, defined by rock formations in the Earth's crust. It will enable navigation in GPS-degraded or -denied environments, such as in the presence of GPS jamming. It is impractical to use ground-based proton-precession/Overhauser magnetometers to obtain the necessary measurements with sufficient spatial resolution and coverage; aerial surveys are required. Lockheed Martin has recently developed its *Dark Ice* technology, which uses a NV-center based vector magnetometer for this purpose (see also section III C). Depending on the flight altitude, these should allow spatial resolution down to $\sim 200$ m, while the small SWaP could allow operation onboard small UAVs [132].

#### *4. Magnetometers in manned aerial vehicles*

Currently, detailed magnetic observations for geological surveys [13, 14], unexploded ordnance detection [133, 134], magnetic anomaly detection (*e.g.*, of submarines or sea mines) [135], and other applications [5, 11] are primarily conducted using magnetometers on manned vehicles, be they land-based [5, 136], aircraft [134, 137, 138], ships [139, 140], or underwater vehicles [141]. Manned aircraft are relatively large and able to generate significantly higher power than UAVs, making them suitable platforms for high-sensitivity airborne magnetometer solutions, such as the SQUID-based tensor magnetic gradient measurement system UXOMAX [142].

SQUID (Superconducting QUantum Interference Device) magnetometers [86–93, 143–148] consist of a superconducting loop split by one or two Josephson junctions [143, 144]. The current circulating in the superconducting loop, and the corresponding voltage drop across the Josephson junction, is sensitive to the magnetic flux threading the loop. SQUID magnetometers offer high magnetic field sensitivity (sub-fT/ $\mathrm{Hz}^{1/2}$ [91, 145]), high dynamic range (they can operate in the Earth's magnetic field [86, 88]), a large range of spatial resolutions (down to the nanometre scale [146, 149]), and broad bandwidth operation (DC to GHz [147]). To achieve superconductivity the SQUID needs to be operated in a cryogenic environment; this incurs large operating costs and is incompatible with low SWaP applications, although they have been used aboard planes and helicopters [145] for applications such as nondestructive archaeology and geomagnetic evaluation [148]. Promising advances are being made towards the creation of ambient-condition superconducting materials, which would lead to improved SWaP requirements for SQUIDS in the future [150], but such materials remain highly speculative and may never eventuate.

#### *5. Magnetometers in unmanned aerial vehicles*

Recent field demonstrations of geomagnetic surveys and magnetic anomaly detection using unmanned aerial vehicles (UAVs) have highlighted several advantages of low SWaP magnetometers for autonomous or remote-controlled surveys [151–155]. UAVs, for instance, allow for exquisite spatial resolution of a few meters, high sensitivity due to low flight altitude, and easy access to rugged terrain [156], while saving cost and operator time. UAV-based surveys are particularly efficient for detecting small targets, such as unexploded ordnance and landmines [151], and are predicted to significantly enhance magnetic mapping capabilities which, in turn, enable improved navigation in GPS-denied environments [157]. However, the most sensitive magnetometer technology with sufficiently low SWaP to be used on typical UAVs is fluxgate magnetometry [158], which has sensitivity several orders of magnitude poorer than techniques such as SQUID magnetometry [159].

Possible alternative high-sensitivity instruments include miniaturized atomic magnetometers [160] and ultra-sensitive integrated magnetostrictive magnetometers [11, 20, 161].

Magnetostrictive magnetometers [11, 20, 162–165] rely on the strain induced in a magnetostrictive material (such as galfenol or Terfenol-D) for detection of the magnitude of applied magnetic fields. Depending on the design of the magnetometer, applying a magnetic field to the magnetostrictive material may cause motion, stress, a force or a torque, which can be detected in a number of ways, but are usually read out electronically or optically. One such magnetometer, using optical readout, uses a magnetostrictive material deposited on an fibre optic interferometer to change the relative path length, and hence the relative phase of laser light, in the two arms of the interferometer [11]. This integrated device has low SWaP (weight $\approx 110$ g and operating power $< 3$ W), a sensitivity of 10 pT/ $\mathrm{Hz}^{1/2}$ over the 1 Hz to 100 Hz frequency range, and a dynamic range of $> 100$ μT (sufficient to operate within the magnetic field at the surface of the Earth). An alternative magnetostrictive magnetometer design uses magnetostrictive material sputter coated onto a microfabricated optomechanical cavity; these optomechanical devices are discussed further in section III.

One of the most common off-the-shelf magnetometers for high field applications are Hall magnetometers [166], which function on the basis of the Hall effect, *i.e.*, an external magnetic field deflects the current flowing through a conductor, leading to a voltage difference perpendicular to the current. Hall magnetometers can measure both AC and DC fields. They are typically used in high field applications and not in precision sensing as they are less accurate than other available magnetometers (peaking around 1 nT/ $\mathrm{Hz}^{1/2}$ [167]). In aerospace, they typically find applications in safety interlocks, rotation gauges, and proximity sensors to ensure safe operation of craft, rather than use as scientific instrumentation.

## III. EMERGING MAGNETOMETERS

Heritage magnetometers – including fluxgate, proton-precession, and optically-pumped magnetometers – have proven utility in space missions, and will continue to be used into the future. However, they have limitations. For instance, FGMs suffer from drifting scale factors and voltage offsets with both time and temperature, requiring periodic recalibration [71]. Proton-precession magnetometers and optically pumped magnetometers exhibit excellent sensitivity (*e.g.* 10–50 pT RMS), absolute accuracy (0.1–1.0 nT), and dynamic range (1–100 μT) [71], but they have considerable mass (> 1 kg), high power requirements (> 10 W), and large volume (> 100 $cm^3$). These 'workhorse' scientific instruments are unsuitable for use in many emerging aerospace applications, particularly in view of the trend towards smaller platforms and probes, *e.g.* CubeSats, NanoSats, PocketQubes, *etc.* To meet the challenges of sensing on small craft, magnetometers must achieve reductions in size, weight, and power ('SWaP') whilst preserving or even enhancing performance.

In general, the magnetometers discussed so far are close to the limits of their applicability. Accordingly, new types of magnetometer need to take their place to go beyond these limits. Some promising candidates are SQUIDs, atomic, NV, MEMS and optomechanical magnetometers. These new sensors also have functionality limits but we are still far from reaching them.

### A. Atomic magnetometers (including SERF)

Atomic magnetometers [18, 69–74, 168, 169] consist of a vapor of alkali atoms (usually K, Rb or Cs) enclosed in a glass cell, generally heated to about 400 K. When a laser beam passes through the vapor cell, the spins of the atoms' unpaired electrons align in the same direction. If a magnetic field is present, the electrons precess, which leads to a polarization or amplitude change in the transmitted light. This can be detected and used to infer the magnetic field. The sensitivities achieved can be very high, on the order of 160 aT $Hz^{1/2}$ [69, 168], with spatial resolution as small as the millimeter scale [69, 70, 73, 74]. Some have a high dynamic range and can operate in the Earth's magnetic field [69, 70], while others have a low dynamic range and require magnetic shielding or close-loop operation [71, 72, 74]. Operation bandwidths typically range from DC to $\sim$ 1 kHz. The atom–light interaction is sensitive to the orientation of the magnetic field, so this type of magnetometer is suitable for vector magnetometry. The most sensitive commercially-available magnetometer is based on atomic magnetometry; these can achieve a sensitivity of 300 fT at 1 Hz [170].

Recent developments in chip-scale atomic magnetometers – such as magnetometers fabricated with silicon micro-machining techniques as part of the "NIST on a Chip" program – have demonstrated a significant reduction in SWaP [18], making them competitive candidates for future CubeSat and UAV projects. The size of these vapor cells is about that of a grain of rice. It is anticipated that such a magnetometer could be placed aboard low cost CubeSats used for detection of the Earth's magnetic field as well as for measuring the magnetic fields of other planets.

Other authors, such as Korth *et al.* [71], have proposed miniaturized atomic scalar magnetometers based on the $^{87}Rb$ isotope for space applications. This magnetometer is based on a vapor cell fabricated using silicon-on-sapphire (SOS) complementary metal-oxide-semiconductor (CMOS) techniques. The vapor cell exhibits a volume of only 1 $mm^3$. The multi-layer SOS-CMOS chip also hosts the Helmholtz coils and additional circuitry required to control the atoms, along with heater coils and thermometers used to adjust the Rb vapor pressure. The overall magnetometer system has a total mass of less than 0.5 kg, consumes less than 1 W of power, and demonstrates a sensitivity of 15 pT/ $Hz^{1/2}$ at 1 Hz. This is comparable with high-sensitivity heritage technologies. Accordingly, these magnetometers address the reduction in SWaP (and potentially cost) without sacrificing performance. They are a viable option for integration in SmallSats for space exploration.

Improved absolute sensitivity can be achieved in atomic vapor cell magnetometers by operating with a dense gas at elevated temperatures. Under these conditions, collisions between the alkali atoms no longer scramble the electronic polarisation, improving the sensor's signal-to-noise ratio. These "Spin Exchange Relaxation-Free" (SERF) devices sacrifice dynamic range for sensitivity; SERFs cannot tolerate the $\sim$ μT fields that are able to be sensed with standard vapor cells. As a result, they require magnetic shielding (which is typically heavy) or active magnetic cancellation (which requires additional control circuitry); they also have increased power requirements because of the elevated temperatures involved. Nevertheless, SERFs are promising candidates for nuclear magnetic resonance sensing [171] – as might be used to detect extraterrestrial organic compounds *in situ* – and biomagnetic sensing. They could also be used to perform magnetocardiography or magnetoencephalography on astronauts for non-invasive health monitoring [172, 173]; for example, the heart produces a field of approximately 10 pT outside the body, whilst the brain produces fields of around 1 pT at the scalp [18].

### B. Optomechanical magnetometers

Optomechanical magnetometers are usually optically- and mechanically-resonant mechanical structures ('cavities') that deform when subjected to a magnetic field [60]. The deformation leads to a change in the optical resonance frequency, which can be detected with extremely high precision. In most optomechanical magnetometers [20, 57, 58, 60, 161, 174, 175], the deformation is due to magnetostrictive coatings or fillings that exert a field-dependent force (much like the aforementioned magnetostrictive magnetometers). Related designs [57, 59, 176–179] respond to the magnetic field gradient *via* the dipole force, or enhance the magnetostrictive response using ferromagnetic resonance [180]. It is notable that rapid progress is occurring in optomechanical sensing, not limited to magnetic fields, but also of other aerospace-relevant stimuli such as temperature [181–183], acoustic vibrations [184, 185], pressure [186], force [187, 188], and acceleration [189–192].

To date, the best field sensitivity demonstrated by an optomechanical magnetometer is 26 pT/ $Hz^{1/2}$ at 10.523 MHz [20]. This is competitive with that of SQUIDs of similar size (approximately 100 μm diameter), but without the requirement for complicated and bulky cryogenics. Furthermore, they do not require magnetic shielding – with typical dynamic ranges being $\sim$ 100 μT – and have low power consumption ($\sim$ 50 μW of optical power). They are often sensitive at frequencies up to 130 MHz, where they are limited by quantum phase noise of the optical readout. At intermediate frequencies, optomechanical magnetometers are limited by thermomechanical noise, and classical laser phase noise becomes dominant below approximately $\sim$ 1 kHz (depending on the light source).

Translational research is being undertaken to integrate these devices into low SWaP packages for a range of in-field applications. The chief challenges at this stage are managing stress in magnetostrictive thin films [161], and reducing or mitigating the effects of low-frequency laser noise. Optomechanical magnetometers will become prime candidates for small orbital platforms – both for scientific and communications purposes [11, 193] – and applications on extraterrestrial rovers or other unmanned vehicles with stringent SWaP requirements.

### C. Magnetometers based on atomic defects in solids

Many crystalline materials host defects (substitutions, vacancies, and combinations thereof) that lead to so-called 'color centers', magnetically-sensitive artificial atoms embedded within the crystal that are addressable by microwave and/or optical fields. Silicon vacancies in silicon carbide [194, 195] have been used to detect magnetic fields in proof-of-concept experiments ($\sim$ 100 nT/ $Hz^{1/2}$ ); however, the best-developed defect-based sensors at the current time use nitrogen–vacancy centers (NV) in diamond [19, 61–68, 196–203].

A negatively charged $NV^-$ defect has a triplet ground state ($^3A_2$), a triplet excited state ($^3E$), and two intermediate singlet states ($^1A$ and $^1E$). The energy separation between the sub-levels in the triplet ground state varies with the magnetic field aligned to the NV quantization axis. When illuminated with green light the defect undergoes photoluminescence at 637 nm; the intensity of the emitted light is higher when the $m_s = 0$ ground state sub-level is populated and exhibits a dip when the population is transferred to the $m_s = \pm 1$ sub-levels. The Zeeman splitting of the ground state, and hence the magnetic field the defect is exposed to, can be measured by using a microwave source to drive the ground state population between the sub-levels and observe the corresponding dip in photon emission [196]. The atomic scale of an NV-defect means that NV-magnetometers naturally have a very high spatial resolution (single-defect magnetometers have been demonstrated by *e.g.*, [67]); this property has been utilized for demonstrating nanoscale imaging of biological samples [197, 198]. The $NV^-$ defects have four possible orientations within the carbon crystal lattice, enabling vector magnetometry techniques to be deployed [199–201]. Sensitivities as good as 0.9 pT/ $Hz^{1/2}$ have been demonstrated in laboratory conditions [19, 65] and operation frequencies vary from DC up to a few gigahertz [61, 63, 64] (with different sensitivities across this range).

This is a relatively new technology, only recently integrated for use outside of the laboratory [68]; as such, there are a limited number of near commercially-available options at present (*e.g.* Lockheed Martin's *Dark Ice*).

## IV. BRIEF SUMMARY

Having introduced the major technologies, we are now in a position to summarise some of the typical performance metrics of existing and emerging magnetometers.

Figure 3 shows the sensitivity of various magnetometers as a function of their typical length scale (linear dimension). As expected, heritage devices (fluxgates, helium magnetometers, proton precession magnetometers, *etc.*) have very good sensitivity, enabling use in many areas, but they tend to be relatively large. Chip-scale electronic devices (AMR/GMR, MEMS, *etc.*) are appreciably smaller and typically exhibit reduced sensitivity. Notably, superconducting magnetometers (SQUIDs) and devices with optical readout (optomechanical, NV diamond, atomic vapor cell,

SERF) are almost universally more sensitive than their conventional/electronic counterparts of comparable sensor size. Furthermore, these emerging technologies are still far from the ultimate performance limits that are enforced by their fundamental noise sources. For example, current optomechanical, NV, and SERF magnetometers are approximately two to three orders of magnitude above their sensitivity limits, as shown in Figure 3; even these limits can potentially be manipulated by leveraging quantum-mechanical effects [58, 204–206]. In contrast, some heritage technologies are already at their physical limits, such as air-core search (induction) coils [207]. Notably, fluxgate magnetometers are not yet at their limit [208].

Typical sensor power requirements are indicated in Figure 4. Note that the colored regions do not consider the power drain incurred by support systems such as electronic processing, cryogenics, heating, *etc.* It is evident that heritage magnetometers have similar total power requirements (solid points in Figure 4) to atomic vapor cells, SERFs and SQUIDs, but the new technologies have an advantage in terms of absolute sensitivity. For applications requiring low power consumption and intermediate sensitivities, optomechanical, NV, and chip-scale electronic sensors are most appropriate. Again, we see that optical readout is an enabling factor for both high-sensitivity devices (AVC, SERF) and low-power devices (optomechanical). The power requirements of NV magnetometers are strongly linked to the light collection efficiency of their readout optics, which are currently low; thus, NV magnetometer power requirements are likely to drop significantly in the future (*e.g.*, [209]).

Finally, we consider the usual operating frequencies of magnetometers, as in Figure 5. Note that these frequency ranges do not indicate the magnetometers' typical operating bandwidths (which are usually much smaller than the ranges given here), nor that any one magnetometer in a category can be tuned across the entire range shown (*e.g.*, SQUID magnetometers come in DC and radio-frequency varieties [144], with high-performance guaranteed only in a specified part of the spectrum). Many types of magnetometer are able to operate down to extremely low frequencies (ELF, 3–30 Hz) or even below, though various low-frequency noise sources tend to lead to reduced sensitivity near DC. As already touched upon, these frequency ranges are important for heliospheric/geomagnetic field mapping, through-water or through-earth communications, *etc.* Conversely, radio frequency operations up to the VHF range ($3 \times 10^8$ Hz, typical of commercial FM radio) are possible with optomechanical and SQUID magnetometers, permitting magnetic antennae for interplanetary or orbital communications, *etc.*

All in all, there is no 'one size fits all' magnetometer technology, nor is there ever likely to be! Careful consideration of the parameters discussed above – plus others that we have not focused on, such as bandwidth, drift, and dynamic range – is required when selecting which magnetometer to deploy for a given application.

## V. CONCLUSIONS

In this review paper we have provided a broad overview of the various aerospace and space-based applications enabled by precision magnetometry. In the context of these applications we first discussed existing magnetometry platforms, highlighting the advantages and limitations of each. No doubt, these 'workhorse' platforms will continue to enjoy use and development well into the future. However, we have identified some emerging aerospace applications, particularly those involving smaller platforms and probes (*e.g.*, CubeSats, NanoSat, PocketQubes, *etc.*), that require magnetometers with lower size, weight, and power ('SWaP') needs, and equivalent or even enhanced performance. Since many of the existing magnetometers are already close to their physical performance limits, new types of magnetometer must be developed to take their place in these emerging applications. We have highlighted some promising magnetometers that are currently under development, alongside the applications that they may enable.

To compare existing and emerging magnetometers, we provided an overview of relevant operational parameters – field sensitivity, power consumption, detection frequency range, sensor size *etc.* – for different classes of magnetometers. Finally, we identified that optical readout is a key route towards improved sensitivity and SWaP for next-generation devices.

## CONTRIBUTIONS

Conceptualization, W.B. and F.M.; literature search and writing, all; surveyed historical missions, B.V. and F.M.; surveyed emerging devices, J.B., H.G., E.B., F.G, S.F, G.I, and W.B.; compiled figures, J.B.; compiled tables, B.V.; project leaders, W.B. and F. M.

**FUNDING**

This research is supported by; the Commonwealth of Australia, as represented by the Defence Science and Technology Group of the Department of Defence (NGTF QT30/40); the Commonwealth of Australia, as represented by the Australian Research Council Centre of Research Excellence for Engineered Quantum Systems (EQUS, Grant No. CE170100009); and the United States of America, as represented by the National Aeronautics and Space Administration (NASA) Space Communications and Navigation (SCaN) program. S.F. acknowledges funding from the EQUS Translational Research Program (TRL, EQUS). E.B. acknowledges funding from an EQUS Deborah Jin Fellowship. F.G. acknowledges funding from Orica Australia, Pty Ltd.

The authors declare no conflict of interest.

---